\documentclass{emulateapj}

\newcommand{\myemail}{pejcha@astronomy.ohio-state.edu}
\newcommand{\actaa}{Acta Astron.}

\newcommand{\tmax}{\ensuremath{t_{\rm max}}}
\shorttitle{Rebrightenings in classical novae}
\shortauthors{Ond\v{r}ej Pejcha}

\begin{document}

\title{The time-dependent rebrightenings in classical nova outbursts: \\ a late-time episodic fuel burning?}

\author{Ond\v{r}ej Pejcha}
\affil{Department of Astronomy, The Ohio State University, 140 West 18th Avenue, Columbus, OH 43210, USA}
\email{\myemail}

\begin{abstract}
A significant fraction of novae exhibit a series of rebrightenings on the decline branch of their light curves. We use visual observations to study this phase in several well-observed novae. We find that these rebrightenings are isolated flare-like events on otherwise smooth light curves and we show that in most novae in our sample the time intervals between consecutive flares gradually increase as a geometric series; rebrightenings are equally spaced in logarithmic time. We also find a correlation between the rate of increase in the time between rebrightenings with the speed class of the nova in the sense that slower novae tend to increase the time intervals faster. We attribute these rebrightenings to instabilities in the envelope hydrogen burning and we mention other cases of such timing pattern in astronomy and natural sciences in general.
\end{abstract}


\section{Introduction}

Decay light curves of explosive phenomena are often bumpy. Optical and X-ray light curves of gamma-ray bursts are superimposed with flares \citep[e.g.][]{stanek07,dai07,burrows05,obrien06} suggesting late-time activity of the central engine. Similar features have been observed during outbursts of X-ray novae \citep[e.g.][]{callanan95,chen97} and of an accretion-driven millisecond pulsar \citep{wijnandsetal01}. A fraction of classical novae also exhibit a series of rebrightenings that were observed for the first time during the 1901 outburst of GK~Per \citep{campbell1903}. These rebrightenings are the subject of this Letter.

An outburst of a classical nova is a result of a thermonuclear runaway in the matter accreted onto the surface of a white dwarf, which ejects an expanding envelope \citep[for a review see e.g.][]{warner95}. Maximum visual brightness occurs when the pseudo-photosphere reaches its maximum radius. Then, the pseudophotosphere shrinks and heats while keeping the bolometric flux essentially constant. This results in a decrease in the visual flux as the bolometric luminosity is radiated at shorter wavelengths. After the ejection the hydrogen remaining on the white dwarf continues burning near the Eddington limit for several weeks to several years \citep[e.g.][]{prialnik86,ogelman93,krautter96,ness07,ness08}.

Roughly a third of very fast or fast novae \citep[p.~259]{warner95} decline smoothly. The rest either show a series of rebrightenings or they create dust which brings the visual flux to a deep minimum. This time period, which usually occurs $3$--$4$ mag below visual maximum, is commonly called the ``transition phase" because of spectroscopic changes toward a nebular spectrum. For simplicity, we will restrict the term ``transition phase" throughout the Letter only to the rebrightenings. 

The underlying physical reason for the transition phase is not yet clear, for a recent review of the proposed models, see \citet{csaketal05}. They spectroscopically monitored the transition phase of V4745~Sgr and found that during the rebrightenings the continuum emission rose and that the spectral lines showed P-Cyg profiles, which suggests a temporary increase in mass-loss rate. On the basis of these observations, \citet{csaketal05} suggested that the rebrightenings are envelope-hydrogen-burning instabilities. They also observed a gradual increase of the time intervals between rebrightenings, which was also noted previously in GK~Per and DK~Lac by \citet{bianchini92}.

In this Letter, we analyze visual light curves of novae that experience the transition phase. We show that these rebrightenings are indeed of eruptive nature and that they follow a specific pattern that relates to the overall rate of decline of the light curve of the nova. We conclude that the rebrightenings are likely caused by hydrogen-burning instabilities and mention other examples of similar timing pattern in astronomy and in the theory of self-organizing critical systems.

\section{Visual light curves}

\begin{figure*}
\plottwo{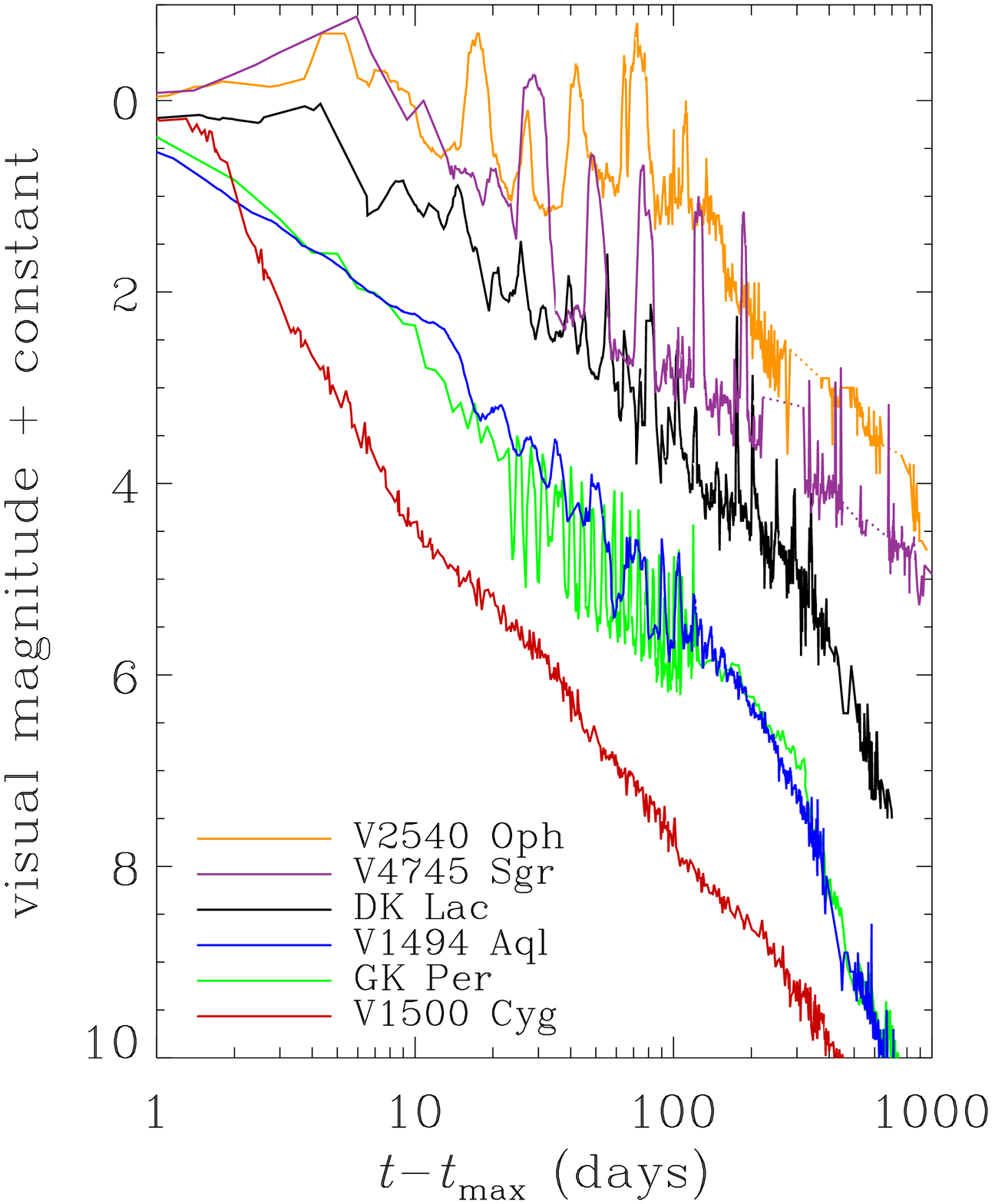}{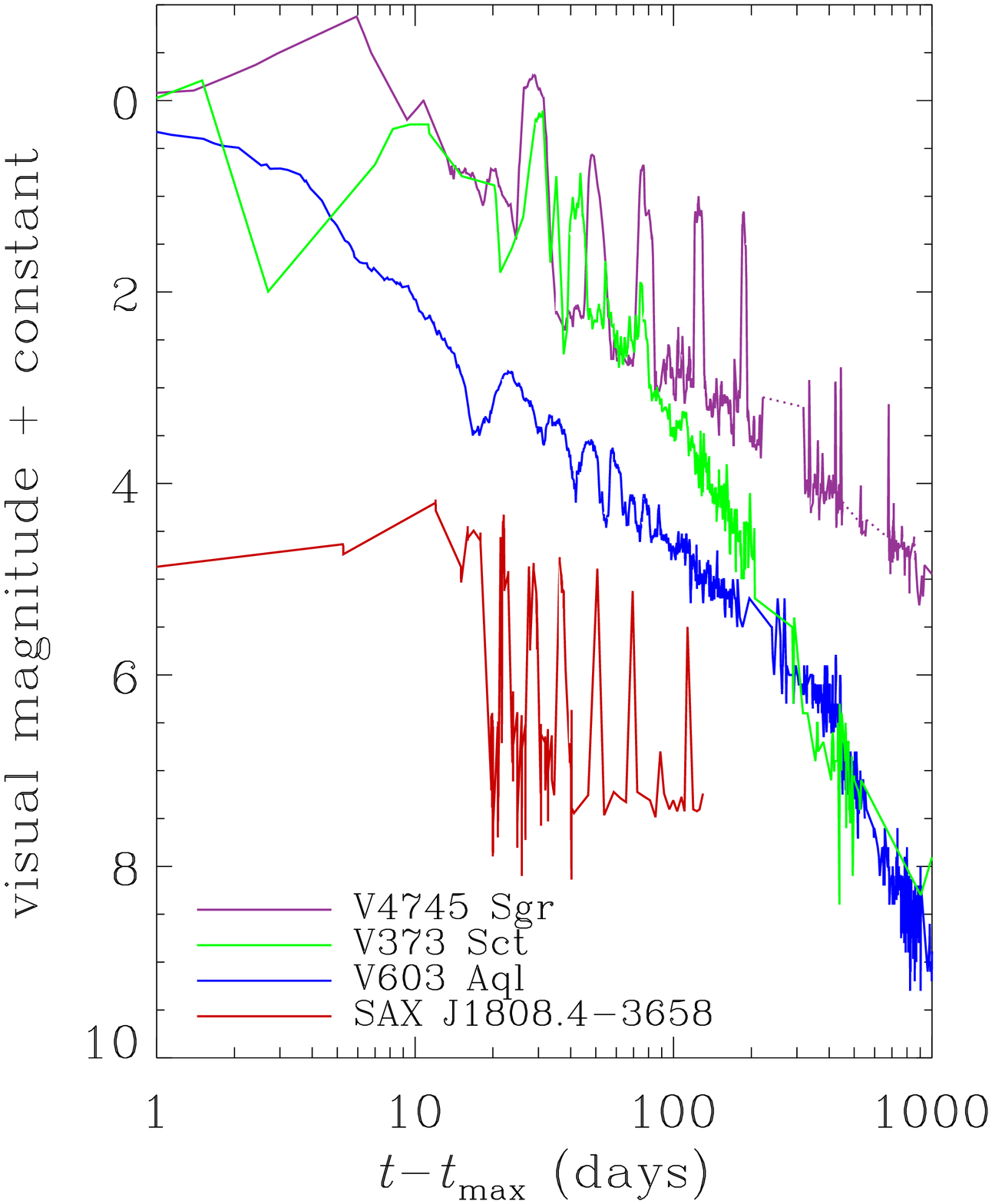}
\caption{Light curves of novae in our sample smoothed using moving averages with bin size of $1$ day. The data were shifted so that the maximum occurred at $\tmax$ with brightness $0$ mag. V4745~Sgr is plotted for reference in both panels. The right panel shows also X-ray light curve of accreting millisecond pulsar SAX J1808.4-3658 \citep{wijnandsetal01}.}
\label{fig:lc}
\end{figure*}

To study the short timescale variations during the transition phase, we need well-sampled light curves obtained using a standard-enough detector so that we can directly compare different objects that erupted during the last $\sim$100 years. Due to the plethora of variable and often strong emission lines, a slight difference in the detector response function could create substantial systematic shift in the magnitude. We choose to use visual observations carried out by amateur astronomers. We obtained magnitude estimates from the archives of the AAVSO\footnote{\url{http://www.aavso.org}} \citep{henden09} and AFOEV\footnote{\url{http://cdsweb.u-strasbg.fr/afoev/}} for V603~Aql, V1494~Aql, DK~Lac, V2540~Oph, and V4745~Sgr. For V4745~Sgr, we supplement these data with visual observations from the VSNET\footnote{\url{http://www.kusastro.kyoto-u.ac.jp/vsnet/}} archive and with the $V$ band measurements of the All Sky Automated Survey \citep[ASAS;][]{pojmanski02}. For GK~Per, we use the mean light curve of \citet{campbell1903} constructed from visual observations. We also include V373~Sct and its visual light curve published by \citet{rosino78} and supplement it with AAVSO and AFOEV observations. To our knowledge, the above mentioned objects represent the best-observed novae exhibiting sufficient number of rebrightenings to study their timing evolution. Note that the error in a single visual estimate is usually assumed to be about $0.3$ mag \citep{kissetal99} and we investigate events with amplitudes typically greater than $1$ mag and defined by many observations of several observers. The rebrightenings we discuss here are robust.

In Figure~\ref{fig:lc}, we show the visual magnitudes of novae in our sample as a function of time. To illustrate a nova light curve with a smooth decline, we include a very fast nova, V1500~Cyg \citep{mattei93}. We plot moving-averaged data with bin size of $1$ day, and we align the light curves so that the visual magnitude at the time of maximum $\tmax$ is $0$ mag. This procedure is potentially problematic in the case of V4745~Sgr and V2540~Oph, because the rebrightenings began shortly after discovery, unlike other novae. However, prediscovery ASAS observations of V4745~Sgr rule out the possibility that the true visual maximum was missed. Given the overall similarity of light curves of both objects, we suggest that the same is true for V2540~Oph and we set $\tmax$ for both objects to be approximately the time of discovery. Importantly, an uncertainty of a couple of days in $\tmax$ does not affect any of our conclusions.

Shortly after maximum, the visual flux of novae decreases as a power-law function of time \citep{hachisu05}, which can be observed as a linear decline in our Figure~\ref{fig:lc}. Several hundred days after $\tmax$, the slope of the decline steepens, which is caused by a decrease in the wind mass-loss rate. The shell hydrogen burning stops shortly afterward \citep{hachisukato06}. The rebrightenings start to occur approximately $20$ days after the maximum. At this time, fast novae like GK~Per or V1494~Aql have faded by about 4 visual magnitudes while the slow novae V2540~Oph and V4745~Sgr are still close to their maximal visual brightness.

\begin{figure}
\plotone{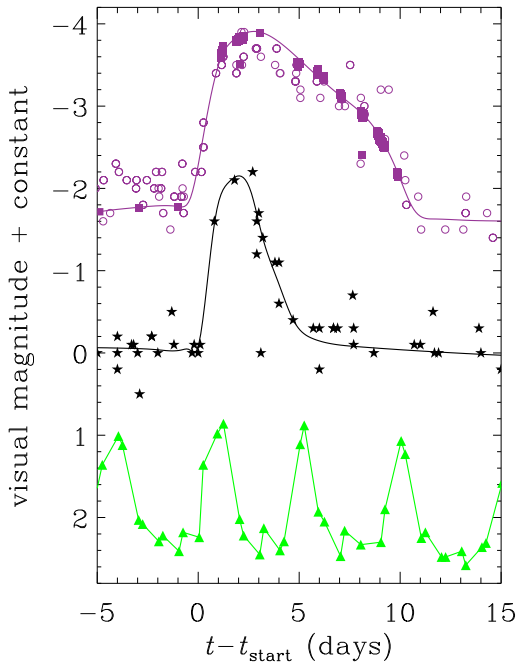}
\caption{Details of individual rebrightenings. Data for GK~Per are plotted with green triangles, for DK~Lac with black stars, and visual data (purple open circles) for V4745~Sgr are plotted along with ASAS $V$-band measurements (purple squares). The light curves were aligned so that the rebrightenings start at $t_{\rm start}$ and arbitrary shifts in magnitude were applied for illustration purposes. Purpose of the solid lines is to guide the eye when reading the plot.}
\label{fig:max}
\end{figure}

To investigate the nature of rebrightenings more closely, we plot in Figure~\ref{fig:max} a blow-up of well-covered rebrightenings for V4745~Sgr, DK~Lac ,and GK~Per. We see that the rise of brightness is much faster than the decline back to the base level. Further, for all novae the amplitude of the rebrightenings ranges from $1.0$ to $2.0$ mag, without any obvious time dependence, although the duration of individual rebrightenings ranges from $2$ days for GK~Per to $10$ days for V4745~Sgr. We note that even for GK~Per, where the time interval between the maxima is about $5$ days, the individual rebrightenings are well separated by a period of essentially constant magnitude. It should be noted, however, that the data for V603~Aql and V1494~Aql do not show the rebrightenings so clearly separated. We attribute this to the lower quality of data available for these stars and to the averaging that smears out details in the light curves. Indeed, the phase of constant brightness stands out more clearly in unbinned data. Given this evidence, we consider the transition phase rebrightenings to be appropriately described as {\em flares} on top of a continuous underlying light curve.
\enlargethispage{\baselineskip}

\section{Long-term behavior of the flares}

\begin{figure*}
\includegraphics[width=\textwidth]{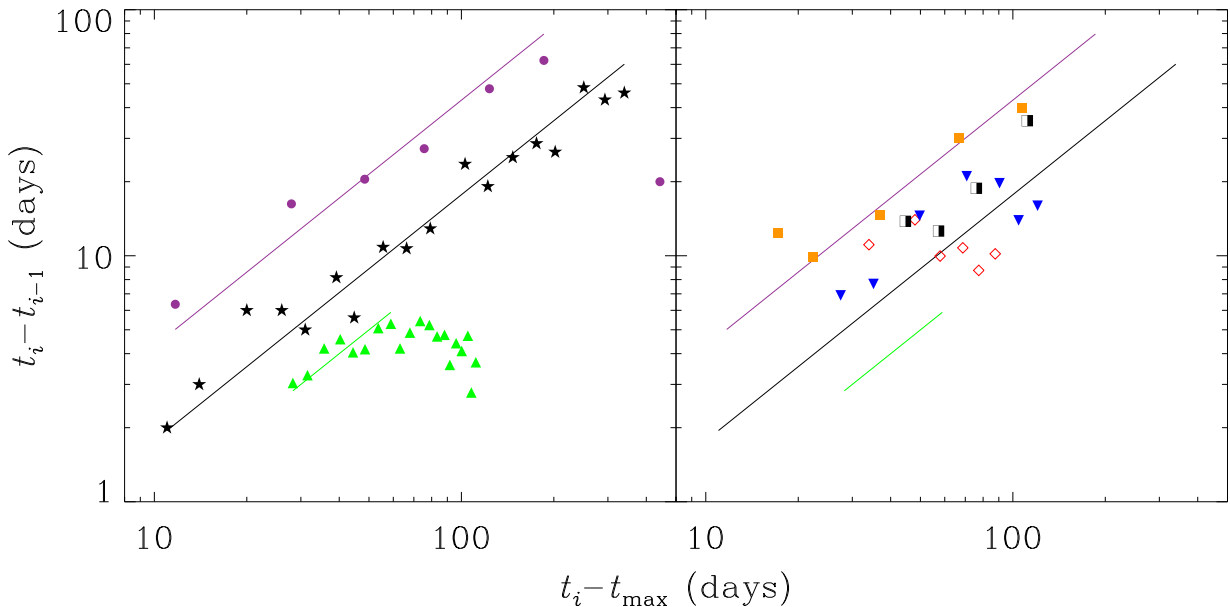}
\caption{Time intervals between two consecutive flares $t_i-t_{i-1}$ as a function of time elapsed since the optical maximum $t_i-\tmax$. Left panel: data for V4745~Sgr (purple circles), DK~Lac (black stars), and GK~Per (green triangles pointing up). Lines mark fit of Equation~(\ref{eq:powlaw}) with $b=1$ to the all points of DK~Lac, first six points of V4745~Sgr and first eight points of GK~Per. Right panel: the same as in the left panel but for V2540~Oph (orange squares), V1494~Aql (blue triangles), V603~Aql (open red diamonds) and V373~Sct (black half-filled squares). Overplotted are fits of V4745~Sgr, DK~Lac and GK~Per from the left panel.}
\label{fig:times1}
\end{figure*}

Looking again at Figure~\ref{fig:lc} and particularly at the light curve of V4745~Sgr (purple line) we see that the flares have approximately constant separation in logarithmic time. This means that the intervals between consecutive maxima get gradually longer, as was noted for this nova by \citet{csaketal05}. V2540~Oph and DK~Lac exhibit apparently the same behavior. To investigate further this gradual increase of separation between flares, we have determined times of maxima $t_i$, $i=1,2,3,\ldots$, by fitting a low-order polynomial to the data with the flare and reading its maximum position. We estimate the error of $t_i$ to be $10\%$ of the time interval used for polynomial fitting. This translates to less than $1$\,day for most of the flares. We experimented with fitting data averaged with different bin sizes and with unbinned data and conclude that the differences in maxima timings $t_i$ are smaller than the assumed error. We compared our maxima timings of DK~Lac with those of \citet{bochnicek51} and found good agreement. Data of \citet{bochnicek51} contain timings of several maxima that occurred prior to those that we were able to derive from our data. We add these timings to ours but we note that our conclusions below are not affected in any way by their inclusion. We also point out that in addition to the seven flares observed by \citet{csaketal05} in V4745~Sgr the ASAS data contain two additional flares that occurred at JD\,$2453165$ and JD\,$2453185$ ($t-\tmax=423$ and $443$\,days) which have lower amplitude and shorter duration than the preceding flares. Two single brighter data points indicate that there might have also been other flares at JD\,$2453085$ and JD\,$2453420$ ($t-\tmax=343$ and $678$\,days) but we cannot show that they are maxima.

In the left panel of Figure~\ref{fig:times1}, we show the time interval between two consecutive flares $t_i-t_{i-1}$ as a function of time elapsed since $\tmax$ for V4745~Sgr, DK~Lac, and GK~Per. We see that the timings for V4745~Sgr and DK~Lac follow a power law
\begin{equation}
\log (t_i -t_{i-1}) = a+b\log(t_i - \tmax)
\label{eq:powlaw}
\end{equation}
with similar values of the power-law index $b$ but with different normalization $a$. A least-squares fit to the data yields $b = 0.88 \pm 0.04$ for DK~Lac and $b=0.79 \pm 0.04$ for V4745~Sgr, assuming a 2-day error of maxima timings. However, maxima timings for these stars can be well fitted with $b$ being held fixed at $b=1$ as is shown in Figure~\ref{fig:times1}. This value of $b$ has a convenient property that the ratio of the times of two consecutive flare maxima is constant: $(t_i-\tmax)/(t_{i-1}-\tmax) = 1/(1-10^a)$. In other words, the times of the flares $t_i-\tmax$ increase as a geometrical series.

The situation is more complicated for GK~Per. Initially, the time interval between the flares increases, as noted by \citet{bianchini92}, and the increase agrees with Equation~(\ref{eq:powlaw}) with $b=1$ (as is shown by a fit to the first eight points in the left panel of Figure~\ref{fig:times1}). Afterward, the flare separation remains steady with a possible decrease at $t_i-\tmax\approx 100$\,days. While this behavior is not seen in DK~Lac, we point out that the two latest flares of V4745~Sgr are separated by only about $20$\,days and suggest that a late decrease of flare separation can occur also in novae that otherwise behave according to Equation~(\ref{eq:powlaw}).

Time intervals between consecutive flares in V4745~Sgr grow faster than in DK~Lac, the normalization $a$ of the former is higher than that of the latter. Flare intervals for GK~Per are consistently lower than for both DK~Lac and V4745~Sgr at the same $t-\tmax$. Is there some correlation between this behavior and other parameters of the novae? Looking at Figure~\ref{fig:lc}, we find that the visual magnitude of GK~Per declined faster than that of DK~Lac which is in turn slower than V4745~Sgr. We propose there is a correlation between the normalization $a$ and the rate of decline of the nova. 

To check this assumption further, we plot in the right panel of Figure~\ref{fig:times1} time intervals between consecutive flares for the remaining novae in our sample plus those flare timings of V373 Sct determined from the light curve given by \citet{rosino78} that are supported by observations. We see that V2540~Oph with basically the same light curve follows almost the same power law as V4745~Sgr, albeit with a lower total count of flares. Similarly, flare times of V373~Sct fall between DK~Lac and V4745~Sgr, which is true also for the light curve. This correlation does not entirely hold for V1494~Aql and V603~Aql. V603~Aql is usually regarded as a very fast nova similar to GK~Per and V1494~Aql. However, while its visual light curve declines initially in the same manner as GK~Per, starting from $t-\tmax \approx 50$ days it declines less steeply and in the final stages of the light curve V603~Aql is slower than GK~Per by a factor of about $2$. On the other hand, the light curve of V1494~Aql is virtually the same as of GK~Per, but the flare separations are significantly larger and until $t-\tmax\approx 90$ days they increase according to Equation~(\ref{eq:powlaw}). Interestingly, the total duration of the transition phase is the same in both objects. 

To put our correlation analysis on more quantitative grounds, we have obtained the rate of the magnitude decline by determining the time-stretching factor $p$ of the light curves of our novae with respect to the light curve of V1500 Cyg, assuming that light curves of novae are homologous \citep{hachisukato06}. We also derived normalizations $a$ while holding $b=1$ fixed for all our novae. Spearman's rank correlation coefficient of quantities $a$ and $p$ is $0.75$ with a $6\%$ chance of random correlation. We conclude that the time intervals between the flares generally increase faster for slower novae, but there exists other parameter or parameters that have also effect as is shown by differences in V1494~Aql and GK~Per.

Finally, we note that while the time interval between the flares increases, the duration of the individual flares does not increase in similar manner. For example, in GK~Per the duration varies randomly by about a factor of $2$ with a mean of $\sim$3 days and in DK~Lac the duration of flares ranges from $4$ to $9$ days. This can be observed also in Figure~\ref{fig:lc} where the flares are becoming increasingly narrower due to the nonlinear abscissa.

\section{Discussion}

During the shell hydrogen burning, when the transition phase occurs, the white dwarf is radiating at its Eddington luminosity \citep{prialnik86}. An isotropic energy release of $10^{43}$--$10^{44}$\,ergs is required to brighten the nova by an additional factor of $\sim$5 for about $5$ days, assuming the rebrightening affects between $3\%$ and $100\%$ of the bolometric luminosity. This can be accomplished either by burning of $10^{-9}$--$10^{-8}$\,M$_\odot$ of hydrogen, by accreting from $10^{-8}$--$10^{-6}$\,M$_\odot$ of matter onto the white dwarf, or by tapping onto the rotational energy of the white dwarf \citep{hachisu09}. We do not consider accretion likely because the white dwarf powers a continuum-driven wind during the shell hydrogen burning \citep{prialnik86}. Rotational energy extraction would happen most likely by employing strong magnetic fields embedded in the expanding envelope and rotating differentially with respect to the white dwarf. When the envelope density decreases enough, the magnetic field restores synchronous rotation driving strong magnetic activities. This process has been suggested by \citet{hachisu09} as an explanation for the secondary maxima of several recent novae. However, a strong magnetic field is not a sufficient condition for the rebrightenings to occur: \citet{schmidt91} found polar strength magnetic fields in a smoothly decaying nova V1500~Cyg. 

\citet{shaviv01} suggested that the transition phase in novae is caused by stagnating super-Eddington winds. \Citet{vanmarle09} performed numerical simulations of porosity-moderated photon-tiring winds and found strong quasi-periodic luminosity variations caused by the fact that a changing fraction of the photon energy is being used to drive the wind. They also found complex patterns of outflowing and infalling matter. We found that the transition phase flares are generally isolated events well separated by intervals of relatively constant or slowly declining visual brightness. Furthermore, \citet{csaketal05} analyzed optical spectra covering the transition phase of V4745~Sgr and did not find any spectroscopic evidence of infalling matter. They concluded that the flares are caused by repetitive instabilities in shell hydrogen burning on the white dwarf. Our estimate of $10^{-9}$--$10^{-8}$\,M$_\odot$ of hydrogen burned is 1--2 orders of magnitude lower than the minimum envelope mass necessary to sustain equilibrium hydrogen burning \citep{sala05}. A late-time increase in the envelope hydrogen burning and the associated mass ejection has also been invoked to explain a secondary peak in V2362~Cyg \citep{lynch08}. Therefore, we conclude that the transition phase flares are likely caused by sudden increases in otherwise steady hydrogen burning in the white dwarf envelope. 

By analysis of flare timings, we have shown that the intervals between consecutive flares increase as a power-law function of time. However, the length of the flares remains essentially constant during the transition phase suggesting that only the trigger of the flare is affected. The power-law exponent is generally the same for all novae, but the normalization is larger for novae with smaller rates of decline. The primary parameter that controls the rate of decline is the mass of the white dwarf: faster novae occur on more massive white dwarfs \citep{prialnik95}. It is then natural to expect that with a more massive white dwarf the intervals between hydrogen-burning flashes will be shorter and also that the flare will last a shorter amount of the time. However, as we have demonstrated on V1494~Aql and GK~Per, which have essentially the same light curve but substantially different intervals between flares, there is one or more parameters that also influence the properties of the flares. We suspect that chemical composition of the white dwarf and its hydrogen-burning envelope, as well as its mass or the mass accretion rate might play a role.

The gradual increase of intervals between individual rebrightenings is an unusual phenomenon in astronomy. We have been able to find only one other example of such behavior: the X-ray and optical light curve of the accretion-driven millisecond pulsar SAX J1808.4--3658 during its 2000 outburst  \citep{wijnandsetal01}. We plot the X-ray light curve in the right panel of Figure~\ref{fig:lc}. However, during the 2002 outburst the flares were more regular, with no apparent increase in time separation \citep{wijnands04}. 

In other areas of physics the phenomenon of logarithmic periodicity occurs as a correction to a power-law behavior and suggests presence of discrete scale invariance \citep[DSI,][p.~276]{sornette04}. A well-known example of DSI is the triadic Cantor set which is geometrically identical to itself only under magnifications equal to powers of 3 \citep{sornette98}. Among examples of DSI and logarithmic periodicity observed in nature we specifically mention the rate of acoustic emissions as a function of time before the rupture of pressure vessels \citep{zhou02a}, and ground-water chemical anomalies preceding the 1995 Kobe earthquake \citep{johansen00,zhou02b}. Theoretical models, which exhibit spontaneous emergence of DSI, include earthquake aftershock simulations based on cellular automata \citep{lee00} and rupture studies with a thermal fuse model \citep{johansen98}. Log-periodic echoing was also found during a critical relativistic gravitational collapse of a massless scalar field \citep{choptuik93,choptuik03}. Finally, we point out that there is no single underlying mechanism for the DSI in general \citep[p.~160]{sornette04}.

\acknowledgments

We are grateful to K.~Z. Stanek, T.~A. Thompson, R.~W. Pogge, C.~S. Kochanek, L.~Bildsten and D.~Sornette for discussions, encouragement, advice on astrophysical processes and critical reading of the manuscript. We thank the anonymous referee for comments that improved the Letter. We acknowledge the observations acquired from the AAVSO International Database, the AFOEV database, operated at CDS, France, and from the VSNET database.

{\it Facility:} \facility{AAVSO}

\clearpage
\end{document}